\documentclass[sigconf]{acmart}

\AtBeginDocument{%
  }

\setcopyright{rightsretained}
\copyrightyear{2026}

\newcommand{\toolNamePlain}{AltIcon}     
\newcommand{\toolName}{\textsc{\toolNamePlain}}

\usepackage{wrapfig}
\usepackage{listings}
\usepackage{tcolorbox}
\usepackage{subcaption}
\tcbuselibrary{listingsutf8}

\begin{document}

\lstset{
    basicstyle=\ttfamily\scriptsize,
    breaklines=true,
    breakindent=0pt,
    frame=single,
    numbers=left,
    numberstyle=\tiny,
    keywordstyle=\color{blue},
    stringstyle=\color{red},
    commentstyle=\color{green!60!black},
    showstringspaces=false,
    captionpos=b,
}

\title{Early Accessibility: Automating Alt-Text Generation for UI Icons During App Development}

\author{Sabrina Haque}
\affiliation{%
  \department{Department of Computer Science and Engineering}
  \institution{University of Texas at Arlington}
  \city{Arlington}
  \state{Texas}
  \country{USA}}
\email{sxh3912@mavs.uta.edu}

\author{Christoph Csallner}
\affiliation{%
  \department{Department of Computer Science and Engineering}
  \institution{University of Texas at Arlington}
  \city{Arlington}
  \state{Texas}
  \country{USA}}
\email{csallner@uta.edu}
  
\begin{abstract}

Alt-text is essential for mobile app accessibility, yet UI icons often lack meaningful descriptions, limiting accessibility for screen reader users. Existing approaches either require extensive labeled datasets, struggle with partial UI contexts, or operate post-development, increasing technical debt. We first conduct a formative study to determine when and how developers prefer to generate icon alt-text. We then explore the \toolName{} approach for generating alt-text for UI icons during development using two fine-tuned models: a text-only large language model that processes extracted UI metadata and a multi-modal model that jointly analyzes icon images and textual context. To improve accuracy, the method extracts relevant UI information from the DOM tree, retrieves in-icon text via OCR, and applies structured prompts for alt-text generation. Our empirical evaluation with the most closely related deep-learning and vision-language models shows that \toolName{} generates alt-text that is of higher quality while not requiring a full-screen input.

\end{abstract}

\setcopyright{none} 
\settopmatter{printacmref=false} 
\renewcommand\footnotetextcopyrightpermission[1]{}

\maketitle

\section{Introduction}

Inferring alt-text for an app icon currently either generates low-quality alt-text or requires programmers to wait until the code of a complete app screen is available. Waiting until a screen is complete causes programmers to lose important context information. Taken together, most programmers today do not use tool support for generating icon alt-text (and we have also observed this lack of tool use in our surveys).

Although federal law such as Section 508 of the U.S. Rehabilitation Act and the Americans with Disabilities Act (ADA) mandates digital accessibility~\cite{ada, 508_act}, most programmers do not add alt-text manually either. Compliance is often overlooked due to the additional manual effort required~\cite{di2022making} and time constraints and lack of prioritization during development~\cite{feng2021auto}. Despite established guidelines, developers often de-prioritize manual accessibility tasks due to tight deadlines, competing feature demands, and a lack of immediate impact on core functionality~\cite{fok2022large, bai2017cost}.

Many apps thus ship with inadequate or missing alt-text~\cite{ross2018examining, alshayban2020accessibility,fok2022large}. Meaningful alt-text however is required by screen readers such as TalkBack and VoiceOver~\cite{talkback,voiceover} to make apps accessible~\cite{chen2021accessible}. To interact with an app, almost all users with disabilities rely on screen readers (e.g., 94\% of users in a recent survey~\cite{screenreader_survey}).

Inferring the meaning of an arbitrary icon remains a fundamentally hard problem. We can only make a limited number of observations, yet have to summarize and generalize these observations to a description that captures the icon's meaning in all possible app usage scenarios. To make this generalization, prior work has leveraged a variety of types of observation (including screenshots, view hierarchy information, and app runtime behavior). We are not aware of work that supports a programmer during development with high-quality alt-texts when the full screen information is not available yet.

At a high level, recent work falls into one of the following two broad categories. In the first category are traditional deep-learning techniques that are specialized for inferring GUI element alt-texts~\cite{chen2020unblind,mehralian2021data}. While they paved the way for subsequent work, these pioneering approaches are trained on relatively small data sets and struggle with generating meaningful alt-text for less-common cases.
In the second category are modern vision-language models (VLMs) such as Pix2Struct~\cite{lee2023pix2struct} and PaliGemma~\cite{beyer2024paligemma}. While these models have been trained on much larger datasets and fine-tuned on widget captioning tasks, they do not perform well on partial screens. These models also rely on extensive UI training data, making them resource-intensive and potentially less adaptable to unseen applications.

To address these challenges, this paper proposes \toolName{}, a method that integrates alt-text generation directly into the IDE, addressing accessibility gaps early during development. This paper introduces a novel application of off-the-shelf large language models to generate context-aware alt-texts for mobile app UI elements without requiring complete screen information or extensive task-specific training. Unlike existing solutions that are more capable of handling issues post-development, \toolName{} is especially beneficial during the early phases of app development, where screen data—such as screenshots, UI elements, view hierarchies, and UI code—are often incomplete. To summarize, this paper makes the following major contributions.

\begin{itemize}
\item This paper presents the first formative study on when and how developers prefer alt-text generation tools to integrate into their workflow.
\item This paper presents \toolName{}, the first method for generating high-quality icon alt-text within the development workflow, enabling a developer to incorporate accessibility early rather than as a post-development fix.
\item This paper empirically compares \toolName{} against the most closely related approaches and its zero-shot GPT-4o baseline on standard metrics and via a user study.
\item This paper analyzes \toolName{} variants' cost-performance trade-offs and the performance contribution of \toolName{}'s components and surveys potential users' perceptions.
\item Our \toolName{} implementation, data, fine-tuning configuration, and evaluation scripts are publicly available~\cite{figshare}.
\end{itemize}

\section{Background}

Android UI development has traditionally relied on XML-based layouts~\cite{layout}, where UI elements are defined in structured resource files and referenced from Java or Kotlin code. Accessibility attributes such as contentDescription are specified directly in XML for components like ImageButton and ImageView, which is the convention we follow for writing this paper.

More recently, Android Jetpack has introduced a fully declarative UI paradigm~\cite{jetpack_compose}, replacing XML with composable functions written in Kotlin. Instead of defining UI elements in resource files, developers construct UIs dynamically using functions like Icon() and Image(), with accessibility metadata such as contentDescription provided within the functions. Despite these differences, the underlying accessibility principles remain unchanged, as screen readers still rely on developer-specified descriptions. \toolName{} can be extended to support Jetpack with minimal modifications, adapting its alt-text generation to declarative UI structures while maintaining the same functionality.

Adding alt-text is often an afterthought in development. Writing meaningful descriptions requires understanding an icon’s intended function, but developers often perceive this task as tedious and time-consuming, resulting in it being deprioritized or ignored~\cite{feng2021auto, di2022making}.

A few years ago a study of a subset of the Rico Android apps dataset~\cite{deka2017rico} has shown that 86\% of clickable ImageViews and 55\% of ImageButtons had an empty or missing contentDescription field, making them inaccessible to screen readers~\cite{ross2018examining}. This accessibility gap still persists today. Our analysis of the recent MUD dataset~\cite{feng2024mud} reveals similar shortcomings, e.g., we found 74\% of Clickable ImageViews and 43\% of ImageButtons still miss a contentDescription.

\subsection{Context-Aware Icon Labeling}

\begin{wrapfigure}{r}{.6\linewidth}
\centering
\includegraphics[width=.48\linewidth]{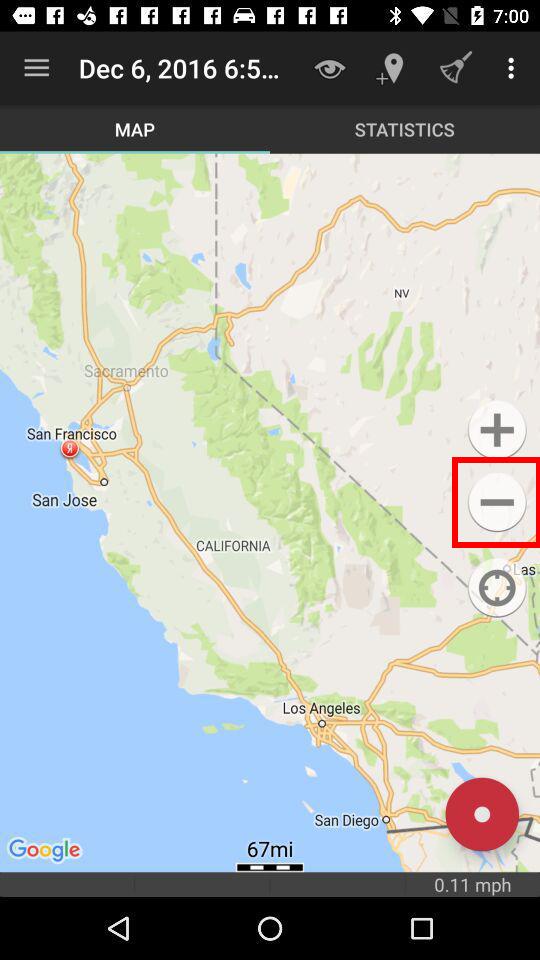}
\includegraphics[width=.48\linewidth]{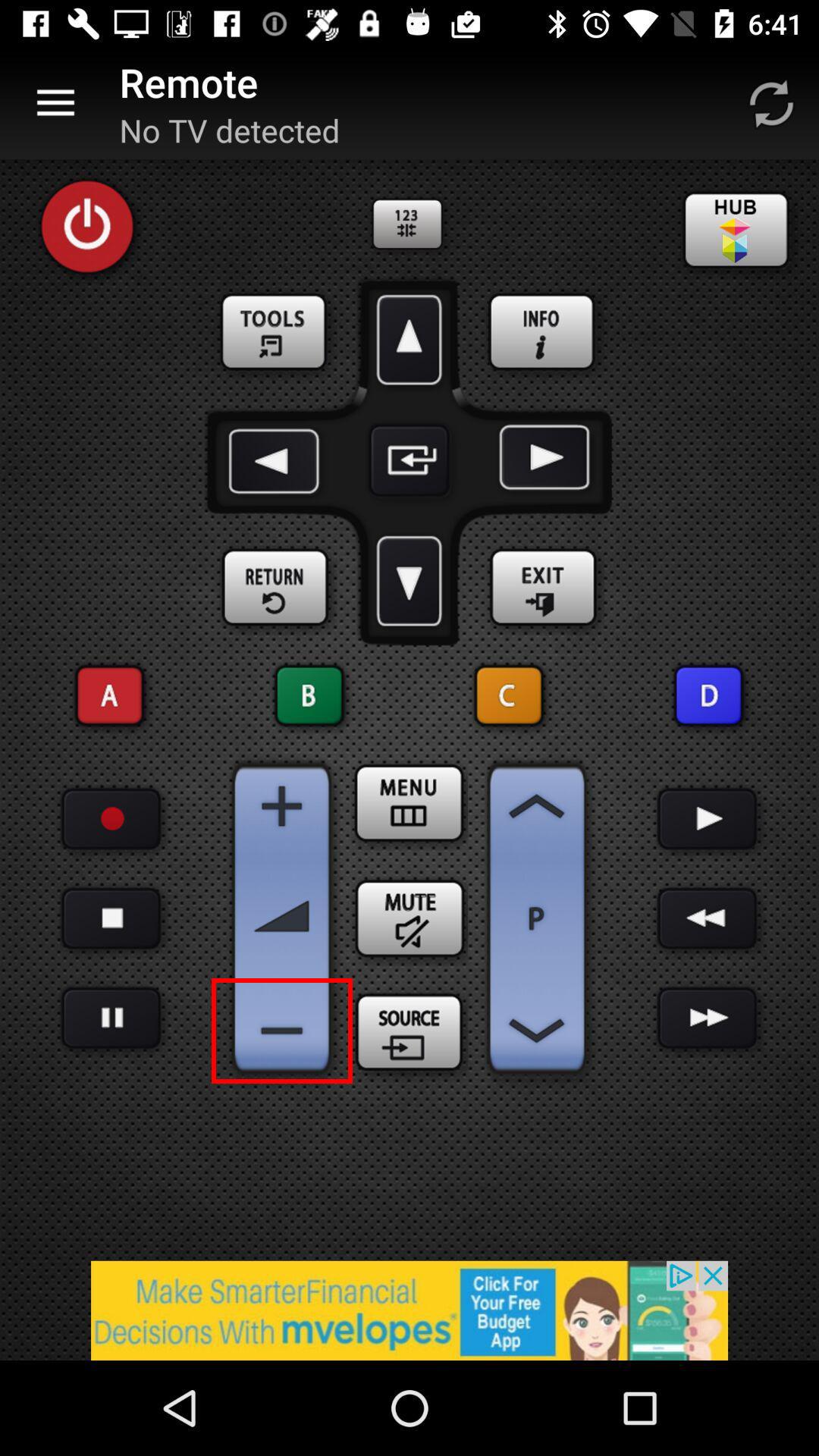}
\Description{Two UI icons are displayed side by side, both featuring a minus symbol. The left icon represents a ``zoom out'' function, and the right icon serves as a ``lower volume'' button.}
\caption{Example zoom out (left) vs. lower volume (right) minus buttons in Rico.}
\label{fig:different_icons}
\end{wrapfigure}

Icons' widespread use in mobile apps can be attributed to their ability to convey information effectively while consuming minimal screen space. Inferring meaningful alt-text for icons is challenging because identical visual representations can serve different functions depending on their context. Without considering surrounding UI elements, automated methods risk generating ambiguous or misleading descriptions. For example, in the Rico dataset~\cite{deka2017rico}, Figure~\ref{fig:different_icons} shows two instances of a ``minus'' icon: one used for zooming out and another for lowering volume. Effective alt-text generation must account for such contextual differences to ensure functionally accurate descriptions.

Early deep-learning models, such as LabelDroid~\cite{chen2020unblind}, treat icons as isolated graphical elements, applying image captioning techniques to generate labels but lack contextual awareness. Coala~\cite{mehralian2021data} improves upon this by incorporating view hierarchy information, allowing for more precise labeling. 

Coala however does not utilize OCR-based in-icon text, limiting its ability to capture any textual hint embedded within icons. 

Both LabelDroid and Coala are deep learning-based approaches, meaning they require large-scale training datasets to achieve strong performance. 

Instead of applying traditional deep learning, recent work adapts VLMs to this task. For example, Pix2Struct leverages the rich visual and textual data from web pages by parsing masked screenshots into simplified HTML, which facilitates the model's ability to integrate visual and textual information during pretraining~\cite{lee2023pix2struct,Donut2022}. The model is trained on 80M screenshots, enabling it to generalize across diverse domains including mobile UI widget captioning. 

Another recent well-known example, the Pali series of vision-language models has analyzed the structural components of UIs~\cite{chen2023pali}. These models are adept at (among others) visually-situated text understanding, which is crucial for comprehensive widget captioning. To efficiently handle noisy image-text data, they use robust pretraining methods, such as contrastive pretraining using SigLIP. 
PaliGemma extends this capability by integrating the SigLIP vision encoder with the Gemma language model, optimizing performance across diverse visual and linguistic tasks~\cite{beyer2024paligemma}. While these models have shown success in UI widget captioning based solely on visual input, they typically process full-screen UIs and are less effective in early-stage development when screens remain incomplete.

\subsection{The Need for Early Accessibility Integration}

Accessibility tasks are frequently postponed to the final stages of development or ignored altogether, making fixes more expensive and less likely to be implemented~\cite{di2022making}. Research highlights the importance of integrating accessibility earlier in the development cycle to reduce technical debt and ensure compliance with accessibility standards~\cite{miranda2022studying}. In agile and iterative development, research calls for accessibility tools to be used ``as early as possible in the projects and as often as possible''~\cite{bai2017cost}.  A recent icon-label longitudinal study has also called for tool support at the moment of change ``At a smaller scale, developer tools might also target moments of change corresponding to new interface elements [..]''~\cite{fok2022large}.

Post development accessibility fixes introduce inefficiencies, increasing costs and technical debt, as developers must revisit previously completed work, disrupting workflow and introducing context-switching overhead~\cite{bi2022accessibility}. Once new features are implemented, revisiting UI elements for alt-text annotation becomes unlikely, leading to permanent omissions. Shift-left accessibility, which advocates for integrating accessibility earlier in the development lifecycle, has been proposed to mitigate these issues~\cite{shift_left}. Additionally, in team-based development, inferring icon functionality while UI screens are incomplete is difficult for developers—especially junior and visually impaired programmers~\cite{feng2021auto}. Adding alt-text early, during development, can help address this issue. 

Most existing accessibility tools operate post-development, detecting violations only after the UI has been implemented. While valuable, they require developers to spend a significant amount of time to localize and fix issues manually~\cite{mehralian2024automated}. Static analysis tools such as Android Lint~\cite{lint} flag missing content description during app development but do not assist in generating meaningful alt-text, leaving the burden on developers. Moreover, developers often lack the awareness to understand accessibility warnings from Lint, sometimes seeking ways to disable them rather than address them~\cite{vendome2019can}. FixAlly integrates accessibility scanners with multi-agent LLM-based suggestions to generate automated repair recommendations for accessibility violations~\cite{mehralian2024automated}. However, these methods only intervene after accessibility gaps have already been introduced. The reactive nature of these tools require developers to address issues long after UI components have been designed. This delay increases technical debt and disrupts development workflows.

\section{Formative Survey}

To understand current practices in alt-text generation for small interactive UI elements (i.e., icons and image buttons), associated challenges, and interest in AI-assisted automation, we conducted an online survey targeting software developers. To get a diverse set of respondents we recruited participants via multiple channels, i.e., outreach to college students with app development experience, Reddit posts (r/androiddev, r/iosdev, r/GooglePlayDeveloper), and direct messages to Android and iOS developers on LinkedIn. To encourage honest feedback the survey was anonymous.

We received 52 responses, which skew toward junior developers (i.e., 60\% selected under one year of mobile app development experience, 19\% 1--3 years, 6\% 4--6 years, and 15\% over six years). As their primary development platform, 60\% selected Android, followed by iOS (13\%), Android+iOS (13\%), web (10\%), and none (4\%). 23\% of respondents stated that they worked on accessible apps (\emph{``Have you worked on apps that require accessibility features (e.g., support for screen readers)?''})

Despite this relatively low number, \textbf{68\% are adding alt-text} to at least some image-based UI elements (\emph{``Do you currently add alt-texts for image-based UI elements (e.g. icons, buttons)?''}). To add these alt-texts \textbf{the most common technique is manual (50\%)} followed by other (12\%) (\emph{``How do you currently generate alt-text for image-based UI elements (e.g., icons, buttons)?''}). The latter includes answers such as using AI-generated suggestions with manual review, relying on third-party teams, and using predefined text.

\begin{table}[h!t]
    \centering
    \begin{tabular}{lr}
        \toprule
        \textbf{Would use plugin to generate alt-text} & \textbf{(\%)} \\
        \midrule
        Yes, regularly & 23\\
        Yes, occasionally & 46\\
        Might use if certain criteria met & 15\\
        Would not use it & 10\\
       
        \midrule
        \textbf{Preferred project stage for use} & \textbf{(\%)} \\
        \midrule
        Wireframing & 19\\
        Screen UI code in-progress & 44\\
        Current screen area UI code done & 17\\
        Current screen UI code done & 4\\
        All screens' UI code done & 12\\
        
        \bottomrule
    \end{tabular}
    \caption{Developer interest and preferred project stage for automated alt-text generation. Omits ``no response'': 6\%, 4\%.}
    \label{tab:alttext_timeline}
\end{table}

\textbf{69\% of respondents would use a plugin to generate alt-text} (regularly or occasionally) with another 15\% conditioned on other criteria (Table~\ref{tab:alttext_timeline}: \emph{``Would you be interested in using a plugin for automated generation of alt-texts in your development workflow?''}). Besides criteria common to plugins, respondents also provided alt-text specific criteria, including \emph{``It would have to identify any text in the image and take that into account when generating alt text''}.

Interestingly only a small minority (16\%) prefers to delay using such a plugin at least until an entire screen is available (Table~\ref{tab:alttext_timeline}: \emph{``If such a tool was available, at which stages of development would you find it most helpful?''}). The option selected most often (44\%) was \emph{``as soon as I add an UI element to the screen code''}, meaning \textbf{developers want to use such a plugin immediately when adding an UI element}. This aligns with what we know about developers' cost of context switching. It is easier to judge the output of such a plugin when working on the corresponding UI element and not having to wait until creating an entire screen UI (or even the entire app UI).

\section{Plugin With 2 Solution Approaches }\label{plugin_architecture}

In our formative study, six out of 10 developers asked for a plugin that generates alt-text while they develop the current app screen. To support this interactive workflow on partially developed screens, \toolName{} runs whenever the developer adds a new icon to the DOM tree and \toolName{} can locate the corresponding icon file in the project directory. For Android development, this means that \toolName{} watches for adding buttons to the app's code (e.g., an ImageButton or ImageView when using Android's traditional XML development style). As prior work has largely focused on Android apps~\cite{deka2017rico,chen2020unblind,mehralian2021data}, we similarly evaluate \toolName{} on Android, but the underlying ideas could also be adapted to other platforms, e.g., iOS.

Generating meaningful alt-text for UI icons requires integrating both visual and textual context information. Since processing images with LLMs is more expensive than processing text, \toolName{} offers two options, trading cost for inference quality. The text-tuned approach, the cheaper option, (\textbf{\toolName{}-TextT}) fine-tunes GPT-4o with extracted textual data only, whereas the more expensive multi-modal fine-tuned approach (\textbf{\toolName{}-MMT}) directly fine-tunes GPT-4o with both textual data and icon images. 

\begin{figure}[h!t]
    \centering
    \includegraphics[width=\linewidth]{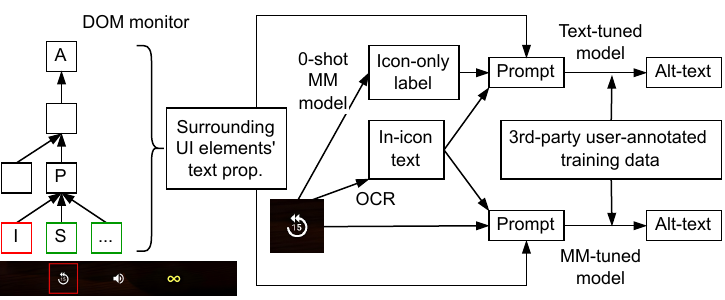}
    \Description{A workflow diagram illustrating \toolName{}'s alt-text generation process. The diagram depicts a UI icon (I) within its functional context, along with its parent (P) and sibling (S) UI elements. Different models generate alt-text for the icon. 
    The ground truth alt-texts include ``go back 15 seconds'', and ``rewind 15 seconds''. The outputs from various models are as follows: \toolName{}-MMT: ``go back 15 seconds''; \toolName{}-TextT: ``go back''; PaliGemma: ``refresh''; Pix2Struct: ``toggle autoplay''.}
\caption{
On adding an icon, \toolName{} extracts icon (I), parent (P), sibling (S), and activity (A) DOM tree info and passes it with icon image-extracted info to one of two fine-tuned models;
ground truth: ``go back 15 seconds'', ``rewind 15 seconds'';
PaliGemma: ``refresh''; 
Pix2Struct: ``toggle autoplay'';
\toolName{}-TextT: ``go back'';
\toolName{}-MMT: ``go back 15 seconds''.}
\label{fig:workflow}
\end{figure}

Figure~\ref{fig:workflow} illustrates the workflow of both \toolName{} options on an icon from the `Abide - Christian Meditation' app. Both \toolName{} options first extract textual icon context information from the DOM tree and in-icon text via OCR. As developers work incrementally, some context information may be missing (e.g., of the icon's sibling nodes). \toolName{} thus does not require complete context information and builds a concise LLM prompt from whatever information is currently available, which then generates alt-text for the new icon.

\subsection{Extracting \& Preprocessing Data}

\toolName{} heuristically detects when the developer adds a clickable icon to the DOM tree based on UI class (currently ImageButton and ImageView). From the DOM tree \toolName{} then extracts the app screen (``activity'') name and the icon’s textual properties (i.e., class name, resource ID, and on-screen text). These strings, crafted by the developer, often contain relevant descriptions, as in the Listing~\ref{lst:icon_context} example.

Since developers often group related UI elements in a container, \toolName{} also identifies the icon's DOM tree parent (container) and extracts similar textual properties from this parent and the parent's direct children---the icon's sibling nodes. Listing~\ref{lst:icon_context} shows the resulting icon context for the Figure~\ref{fig:workflow} example icon.

\begin{lstlisting}[caption={The Figure~\ref{fig:workflow} rewind icon's context \toolName{} extracts from the DOM tree.}, label={lst:icon_context}, numbers=none, float]
{
  "app_activity_name": "is.abide.ui.PlayerActivity",
  "UI_element_info": {
    "class_name": "AppCompatImageButton",
    "resource_id": "rewind_button"
  },
  "parent_node": [
    { "resource_id": "toggle_layout" },
    { "class": "LinearLayout" }
  ],
  "sibling_nodes": [
    {
      "resource_id": "background_music_button",
      "class": "AppCompatImageButton"
    },
    {
      "resource_id": "autoplay_button",
      "class": "ToggleButton"
    }
  ]
}
\end{lstlisting}

If certain properties (e.g., on-screen text) are missing or blank in the DOM tree, \toolName{} just skips them. Currently \toolName{} does not retrieve any developer provided alt-text annotations (Android contentDescription) from the DOM tree. Adding such annotations may improve \toolName{}'s performance.

\subsubsection{In-Icon Text via OCR}

Icons often contain text that conveys functional information. Such in-icon text is typically missing from the DOM tree since it is not stored in the icon’s text property. \toolName{} extracts such in-icon text from the icon file via optical character recognition (OCR) using EasyOCR 1.7.1~\cite{easy_ocr}.

\begin{wrapfigure}{R}{.35\linewidth}
     \centering
    \begin{minipage}[b]{0.45\linewidth}
        \centering
        \includegraphics[width=\linewidth]{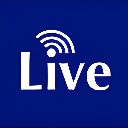}        
    \end{minipage}
    \hfill
    \begin{minipage}[b]{0.45\linewidth}
        \centering
        \includegraphics[width=\linewidth]{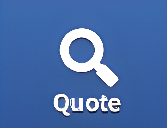}       
    \end{minipage}
    \Description{Two examples of UI icons containing embedded text extracted using OCR from the Rico dataset. The first icon contains the word "Live," and the second icon contains the word "Quote."}
    \caption{In-icon text examples from Rico.     
    }
    \label{fig:ocr}   
\end{wrapfigure}

Figure~\ref{fig:ocr} shows two example icons containing text that would be helpful for alt-text generation but is not available in the DOM tree. \toolName{} successfully extracts the texts (``Live'' and ``Quote'') and adds them to the respective icon contexts.

\subsubsection{Icon-only Label via 0-shot MM Model for \toolName{}-TextT}

We assume the icon's look intends to convey the icon's meaning. We thus fine-tuned \toolName{}-MMT on both icon image files and textual icon context. As such multi-modal fine-tuning is currently more expensive than text-only fine-tuning, we provide the cheaper \toolName{}-TextT option that replaces multi-modal fine-tuning with label-inference followed by text-only fine-tuning.

To infer a label from an icon image we initially trained a deep learning classifier (EfficientNet~\cite{tan2019efficientnet}) on a curated set of 100 icon classes from Flaticon~\cite{flaticon}, the Noun Project~\cite{noun_project}, and LabelDroid~\cite{chen2020unblind}. While this classifier achieved 98\% training accuracy and 94.4\% test accuracy, it struggled with long-tail, less common icons.

As in our experiments, modern models such as GPT-4o performed better even without our training or fine-tuning (``0-shot''), we now use GPT-4o for label-inference. \toolName{}-TextT thus first prompts GPT-4o with the query: \textit{``You are an image classifier. What is the class of this UI icon? Only provide the class as response.''} GPT-4o typically returns a one- or two-word description, which serves as the icon-only label in \toolName{}'s text-based fine-tuning pipeline.

\subsection{Adding Data to Prompt Templates}

\begin{lstlisting}[caption={Prompt template of \toolName{}-TextT.}, label={lst:prompt_TextT}, numbers=none]
"You are an accessibility assistant to a mobile app Developer. A mobile app UI element that looks like an icon with tag '{icon-only label}' has view hierarchy content as below: 
{icon context}
Generate a short (within 2-7 words), DESCRIPTIVE alt-text for the UI element. Provide only the alt-text as output, nothing else. Describe the element as if you were the app developer to HELP VISION-IMPAIRED USER understand its FUNCTIONALITY and PURPOSE. Avoid generic words like 'button', 'image', 'icon' etc."
\end{lstlisting}

From the extracted textual information \toolName{} fills a prompt template for \toolName{}-TextT (Listing~\ref{lst:prompt_TextT}) or \toolName{}-MMT (Listing~\ref{lst:prompt_MMT}). Both templates explicitly define \toolName{}'s accessibility objective.
    
\begin{lstlisting}[caption={Prompt template of \toolName{}-MMT.}, label={lst:prompt_MMT}, numbers=none]
"The attached image is of a UI element (icon) which has the following view hierarchy content: 
{icon context}  
Generate a short (within 2-7 words), DESCRIPTIVE alt-text for the UI element. Provide only the alt-text as output, nothing else. Describe the element as if you were the app developer to HELP VISION-IMPAIRED USERS understand its FUNCTIONALITY and PURPOSE. Avoid generic words like 'button', 'image', 'icon' etc."
\end{lstlisting}

\subsection{Options to Surface Inferred Icon Alt-text}

Existing accessibility tools often require developers to revisit and revise app screens and their UI elements long after development, forcing developers to ``context switch'' to again study the details of such previously developed screens. Such human context switching wastes time and and reduces adoption.

As six of 10 developers in our formative study would like alt-text generation while they develop the current app screen, \toolName{} just injects its inferred icon alt-text directly into the screen's XML layout file when the alt-text is available from the fine-tuned model. Currently, \toolName{} receives this alt-text via a (remote) API call, which may take seconds. Future \toolName{} versions will explore using specialized on-device models to allow fast alt-text injection via an interactive ``syntax-completion'' style.

\toolName{} can also be used programmatically, e.g., in a batch style, to inject alt-text for each icon in an existing app screen layout file. More interestingly, \toolName{} may be used by other IDE plugins. For example, linters such as Android Lint~\cite{lint} are widely used to warn developers about missing or insufficient icon labels. Via calling \toolName{}, a future linter may also suggest a potential fix.

While this \toolName{} evaluation focuses on Android's traditional declarative layout definition style via XML files, the underlying ideas similarly apply to related Android programming styles (e.g., the more recent Android Jetpack) and related platforms.

\section{Evaluation}

To compare \toolName{} with the state of the art, we have first implemented an \toolName{} Rico prototype that operates on data in the Rico format~\cite{deka2017rico}. The Rico dataset is widely used in tool evaluations and makes it easier to put the quality of \toolName{}-inferred alt-texts into context. We measure how well the generated descriptions align with 3rd-party human-written (``ground truth'') alt-text. We have also conducted a user survey on a separate \toolName{} Android Studio plugin prototype that works on Android projects.
We explore the following research questions.
\begin{description}
    \item[RQ1] How does \toolName{} compare with its 0-shot GPT-4o baseline and state-of-the-art approaches (deep-learning, vision-transformer, VLM) on standard automated metrics?
    \item[RQ2] What is the trade-off between fine-tuning \toolName{} on text-only vs. on multi-modal icon input?
    \item[RQ3] How do \toolName{}'s components affect its performance?
    \item [RQ4] How do human evaluators rate the alt-texts generated by the approaches scoring highest on automated metrics?
    \item[RQ5] How do potential users rate the \toolName{} plugin?
\end{description}

We explore RQ1--4 via a subset of WC20, a widely used publicly available dataset for UI widget captioning~\cite{li2020widget}. WC20 extends the foundational Rico dataset~\cite{deka2017rico}, by adding third-party human-annotated alt-text for UI elements, including icons, across 22k mobile UI screens. In these 22k~screens, WC20 annotated a total of 61k~UI elements with up to 3 annotations each by distinct human users, yielding 163k 3rd-party human-produced captions. While other datasets, such as LabelDroid~\cite{chen2020unblind}, rely on developer-provided labels that are often generic~\cite{mehralian2021data}, WC20 offers more diverse, human-written descriptions. Each WC20 screen contains both a screenshot and a corresponding view hierarchy (aka a JSON DOM tree).   

\begin{table}[h]
    \centering
    \caption{WC20 (top) has up to three 3rd-party human labels per widget, our icon label subsets keep one label per training/validation icon at random (r\textsubscript{1}).}
    \begin{tabular}{lrrrr}
        \toprule
        Ground truth & Train & Valid. & Test & Total \\
        \midrule
        WC20 widgets & 52,178 & 4,559 & 4,548 & 61,285 \\
        WC20 labels  &138,342 &12,275 &12,242  &162,859 \\
        \midrule
        WC20 icons             & 18,176 & 1,503 & 1,635 & 21,314 \\
        WC20 icon labels (all) & 48,528 & 4,084 & 4,419 & 57,301 \\
        WC20 icon labels (r\textsubscript{1})  & 18,176 & 1,503 & 4,419 & 24,098 \\
        \midrule
        \toolName{} icons  & 1,425 & - & 1,635 & 3,060 \\
        \toolName{} labels (r\textsubscript{1}) & 1,425 & - & 4,419 & 5,844 \\
        \bottomrule
    \end{tabular}
    \label{tab:WC20}
\end{table}

To create an icon-focused dataset, we filter WC20 to retain only ImageButton and ImageView elements. Of the 22k human-annotated screens, 17k are available (via Rico), containing 26k icons (ImageButton or ImageView). Following prior work~\cite{mehralian2021data}, we remove abnormally large or narrow elements, yielding 21,314 icons across 8,201 screens. For the training and validation sets we randomly select one annotation per icon, yielding 24,098 total labels. Our dataset splits (Table~\ref{tab:WC20}) are consistent with WC20.

\begin{table*}[h!t]
\centering
\caption{DOM \& pixel input (full/partial screen, container, icon) for inference (f,p,c,i), known costs (in USD), and alt-text generation performance of \toolName{} vs. deep learning (Coala, LabelDroid), vision-transformer (Pix2Struct), and VLM (PaliGemma);
bold~=~best score;
underlined~=~runner-up.
\toolName{} is SOTA---and especially for partial screens by a large margin.
}
\begin{tabular}{lccrrrrrrrr}
\toprule
Model & DOM & Pix & Fine-tune & Infer & BLEU-1 & BLEU-2 & ROUGE & METEOR & CIDEr & SPICE \\
\midrule
LabelDroid & -- & i &&& 28.4 & 17.7 & 31.7 & 12.5 & 59.4 & 10.6 \\
Coala & p & i &&& 35.3 & 24.1 & 34.0 & 14.2 & 66.3 & 9.4 \\
\midrule
Pix2Struct-large\textsubscript{f} & -- & f &&& 41.9 & 28.1 & 41.7 & 18.4 & 89.5 & 14.8 \\
Pix2Struct-large\textsubscript{c} & -- & c &&& 35.6 & 23.9 & 35.1 & 15.7 & 73.7 & 10.7 \\
Pix2Struct-large\textsubscript{i} & -- & i &&& 24.1 & 15.9 & 23.1 & 10.6 & 46.4 & 5.0 \\
PaliGemma-448\textsubscript{f} & -- & f &&& 55.4 & 41.6 & 56.2 & 24.7 & 127.1 & 21.8 \\
PaliGemma-448\textsubscript{c} & -- & c &&& 53.6 & 39.3 & 55.9 & 24.1 & 123.5 & 20.6\\      
PaliGemma-448\textsubscript{i} & -- & i &&& 44.0 & 29.2 & 45.3 & 19.2 & 101.3 & 14.3 \\
\midrule
GPT-4o-0-shot\textsubscript{c} & p & c & 0.00 & 0.69 & 35.5 & 18.3 & 38.8 & 19.6 & 66.3 & 14.1 \\  
GPT-4o-0-shot\textsubscript{i} & p & i & 0.00 & 0.68 & 34.1 & 17.4 & 37.4 & 18.7 & 61.6 & 11.9 \\
\toolName{}-TextT & p & i & 29.03 & 1.51 & \underline{60.6} & 43.8 & 57.0 & 26.7 & 134.3 & 22.3 \\
\toolName{}-MMT\textsubscript{c} & p & c & 38.79 & 2.37 & 59.5 & \textbf{44.7} & \underline{59.0} & \underline{26.9} & \underline{136.5} & \underline{22.7}\\ 
\toolName{}-MMT\textsubscript{i} & p & i & 37.33 & 2.05 & \textbf{61.0} & \underline{44.5} & \textbf{59.2} & \textbf{27.7} & \textbf{138.3} & \textbf{23.2} \\
\bottomrule
\end{tabular}
\label{tab:results_metrics}
\end{table*}

\subsection{Experimental Setup}

We empirically compare \toolName{} with representative tools from the most closely related approaches. To make this comparison as fair as possible, we evaluated all tools on the same 1,635~WC20-derived icons and their 4,419 human-annotated labels (Table~\ref{tab:WC20}).

We first reproduce the original results of the deep-learning approaches, using the LabelDroid\footnote{\url{https://github.com/chenjshnn/LabelDroid}, accessed March 2025} and Coala\footnote{\url{https://github.com/fmehralian/COALA}, accessed March 2025} authors' code and configuration settings (e.g., no-attention mode for Coala~\cite{coala2021zenodo}) to train and test these models on their datasets (incl. their DOM-extracted labels). The details of this experiment are part of the \toolName{} replication package on Figshare~\cite{figshare}.

After ensuring reproducibility, we trained both models on our WC20-based icon dataset, using the Widget Captioning paper's exact train-test split (Table~\ref{tab:WC20} ``WC20 icons'' \& ``WC20 icon labels (r\textsubscript{1})'' rows). Specifically, we ran both models' pipelines to extract icons directly from our screenshots. Similar to our work, Coala uses both icon image and its DOM context. We thus convert our DOM from JSON to Coala's XML (incl. folders based on app packages). Compared to reproducing their experiments, LabelDroid and Coala scores differ on our WC20-based data (Table~\ref{tab:results_metrics}), e.g., Coala's CIDEr goes from 134 (for the ``non-predefined'' icons, i.e., excluding the Android standard SDK icons) to 66. This is likely due to Coala's developer-provided labels being more generic vs. the 3rd-party-labeled WC20 providing more informative and diverse labels.

The vision-transformer and VLM approaches are trained on a variety of images and then fine-tuned for specific tasks. We use their best-performing configurations (e.g., 448×448 pixel input size for PaliGemma) and follow their experiment descriptions (i.e., marking the target UI element with a red bounding box and for PaliGemma aspect-ratio preserving input resizing with padding and using prompt ``caption en''). We use the HuggingFace models Pix2Struct\footnote{\url{https://huggingface.co/google/pix2struct-widget-captioning-large}, accessed March 2025} and PaliGemma\footnote{\url{https://huggingface.co/google/paligemma-3b-ft-widgetcap-448}, accessed March 2025} already fine-tuned on over 40k WC20 UI elements. Besides their original full-screen use we also simulate our partial-screen use by blanking out all other UI elements. For partial screens we try both icon-only and icon within its parent container.
Our full-screen results (Table~\ref{tab:results_metrics}) differ from their original results (e.g., bringing PaliGemma's CIDEr score from 148 to 127), as they were on full WC20 (with its many non-icon UI elements).

Modern multi-modal models can be fine-tuned on small sample sets. For \toolName{} we selected 99 common icon classes from Liu et al.~\cite{liu2018learning} and added an "other" category for unmatched icons. For fine-tuning we sampled 1,425 of 18,176 training icons (up to 15 per class). From a WC20 screen's JSON-based DOM tree we infer bounding boxes to crop icons, yielding icons of various resolutions. As a low resolution may hurt icon labeling, we apply Super-Resolution (i.e., Real-ESRGAN~\cite{wang2021real}), standardizing all icons to 128×128 pixels. We fine-tuned \toolName{}-TextT on one randomly selected WC20 label per icon, UI metadata extracted from the layout hierarchy, OCR-extracted in-icon text, and zero-shot inferred icon-only labels. We fine-tuned \toolName{}-MMT on the same data but replacing the icon-only label with the icon image file.

As \toolName{} encodes the icon's container and sibling DOM node information in its prompt we wanted to check if replacing the icon image with the entire container image further improves \toolName{} performance. We thus added a \toolName{}-MMT (and baseline) variant that is trained (and tested) using such larger images, marking the icon with a red bounding box. We may support this MMT\textsubscript{c} variant in the plugin, by taking a screenshot of the IDE's screen preview and parsing its runtime DOM. For the WC20-based evaluation we obtain the container's bounding box directly from the JSON DOM.

We conducted all training and evaluation on a 16GB RAM Intel Core i7-11700F machine with an NVIDIA GeForce RTX 3060 Ti GPU. For fine-tuning (and zero-shot prompting) both \toolName{}-TextT and \toolName{}-MMT use gpt-4o-2024-08-06 via the OpenAI API~\cite{openai,openai_finetuning} under identical conditions over 3 epochs.

\subsubsection{Automated Alt-text Quality Metrics}

We evaluate model outputs against 1,635 ground-truth test pairs of icon (within a screen) plus up to three human-written icon alt-texts. We measure alignment with these human-generated descriptions with the following standard text distance metrics from MS COCO~\cite{ms_coco}, where higher scores mean closer matches.

BLEU~\cite{papineni2002bleu} measures n-gram precision, counting exact matches of word sequences. As icon labels are often short, we focus on BLEU-1 and BLEU-2 (1-gram and 2-gram matches). However, BLEU penalizes spelling differences, synonyms, and paraphrases. ROUGE~\cite{lin2004rouge} instead prioritizes recall, assessing how much information in the reference texts the candidate text captures, but still penalizes synonyms and paraphrases. METEOR~\cite{banerjee2005meteor} improves upon BLEU and ROUGE by including stemming and synonyms matching based on WordNet~\cite{miller1995wordnet} but may still misjudge semantically related phrases.

CIDEr~\cite{vedantam2015cider} assigns higher weight to rare but informative n-grams using TF-IDF weighting~\cite{robertson2004understanding}, making it effective for distinguishing between generic and specific icon descriptions. Its reliance on TF-IDF may sometimes overweight unimportant details, leading to ineffective evaluations~\cite{re-evaluating_captioning_metrics}. SPICE~\cite{anderson2016spice} assesses semantic accuracy by analyzing sentence structure and meaning using scene graphs but may struggle with short alt-texts common in UI labels.
Given their stronger alignment with human judgments~\cite{anderson2016spice}, CIDEr and SPICE are the most suitable metrics for our task.

\subsection{RQ1: SOTA on Automated Metrics}

At a high level, \toolName{} performs better than its GPT-4o zero-shot baseline and its most closely-related approaches across categories and metrics (Table~\ref{tab:results_metrics}). Deep-learning models overall had the lowest scores. Specifically, LabelDroid takes an image captioning approach and lacks contextual understanding, often yielding generic labels (59 CIDEr). Coala improves upon LabelDroid by incorporating partial DOM context (66 CIDEr).

It is interesting that the deep-learning models (after being trained on WC20) have similar CIDEr scores as our GPT-4o baselines out of the box (0-shot). The baselines' SPICE scores are however higher than for the deep-learning models, indicating that (by leaning on its large-scale training) 0-shot GPT-4o can better capture fine-grained semantic relationships out of the box.

The fine-tuned vision-transformers and especially VLMs (CIDEr scores 74 and 122) are competitive with our GPT-4o baselines. This confirms that fine-tuning on a small, task-specific dataset can improve UI accessibility task performance even when comparing with the much larger GPT-4o model.

When comparing the variants of the existing approaches among each other, a common trend is that (0-shot) GPT-4o, Pix2Struct, and PaliGemma scores all drop when providing smaller image subsets, i.e., when going from full-screen to icon-within-container and finally to icon-only. To some degree this is expected as it removes context information, but it is also a challenge when running the existing tools on partial screens.

Finally, all \toolName{} variants achieve higher scores across all metrics than all of the exiting approaches. This is especially notable as the highest-performing existing model variants (Pix2Struct-large\textsubscript{f} and PaliGemma-448\textsubscript{f}) have access to the full app screen, whereas \toolName{} works on just partial screens.

\subsection{RQ2: Text-only vs. Multi-modal Fine-tuning}

To assess the impact of adding image-based features to fine-tuning we compare \toolName{}-TextT (text-only fine-tuning) with \toolName{}-MMT (multi-modal fine-tuning). As a baseline on our 1,635 icons, 0-shot GPT-4o requires zero fine-tuning and low test cost (some USD 0.70).

In exchange for additional fine-tuning plus at least double the test cost, all \toolName{} variants yield much better metric scores than the baseline. Among each other, the \toolName{} variants yield similar metric results. \toolName{}-MMT performs better than \toolName{}-TextT, in exchange for about 1/3 higher fine-tuning and test costs.

\begin{figure*}[h!t]
\centering
\subcaptionbox{
Mathematics:
``delete'', ``delete button'', ``its a standard button for deleting text'' (ref);
``go back'' (PG);
``delete'' (\toolName{}).
\label{fig:screen_good_calc}}{
    \includegraphics[width=0.23\textwidth]{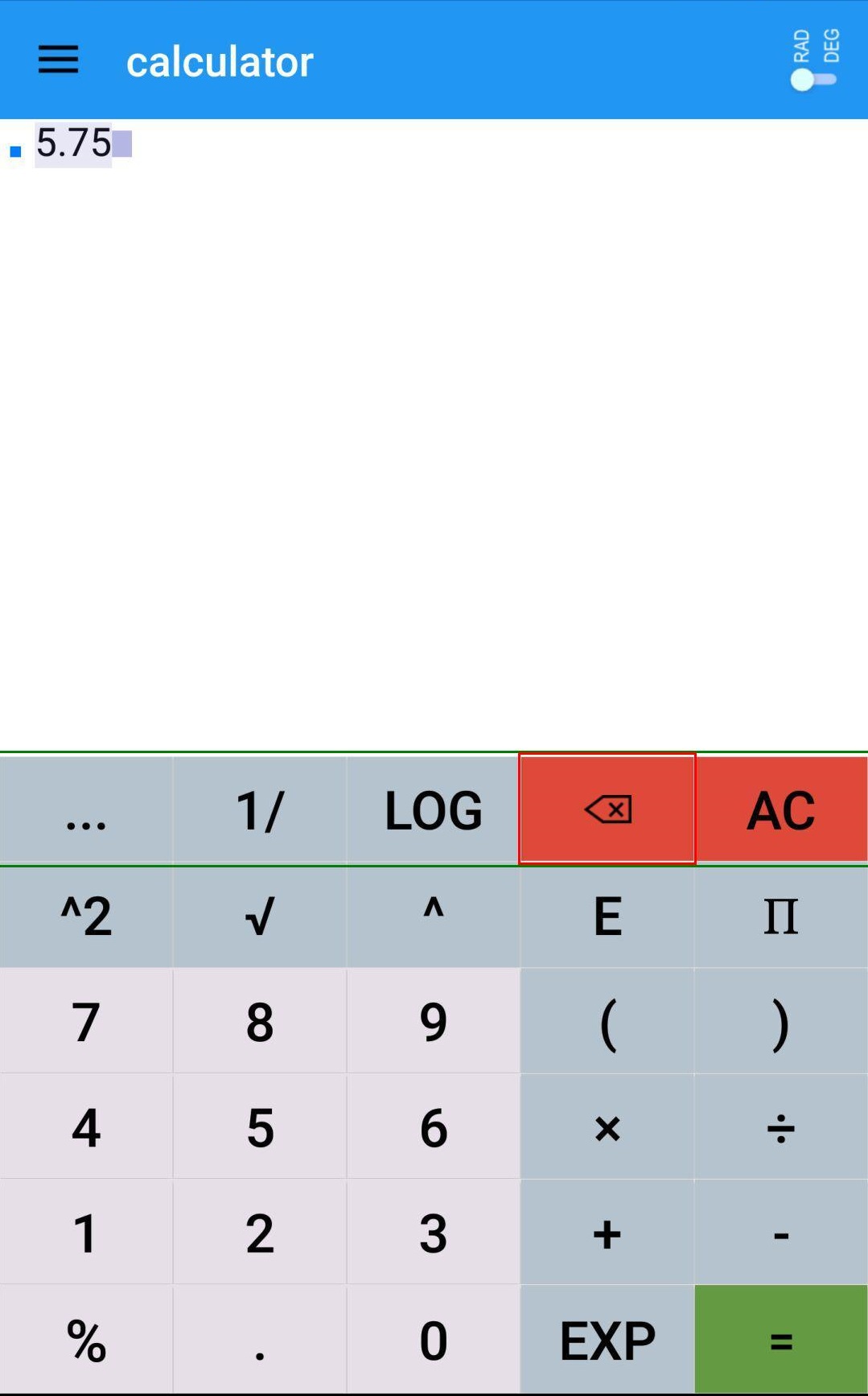}}
\hfill
\subcaptionbox{
ROMEO - Gay Social Network:
``hide profile'', ``turn of camera'' (ref);
``start something'' (PG);
``hide profile'' (\toolName{}).
\label{fig:screen_good_romeo}}{
    \includegraphics[width=0.23\textwidth]{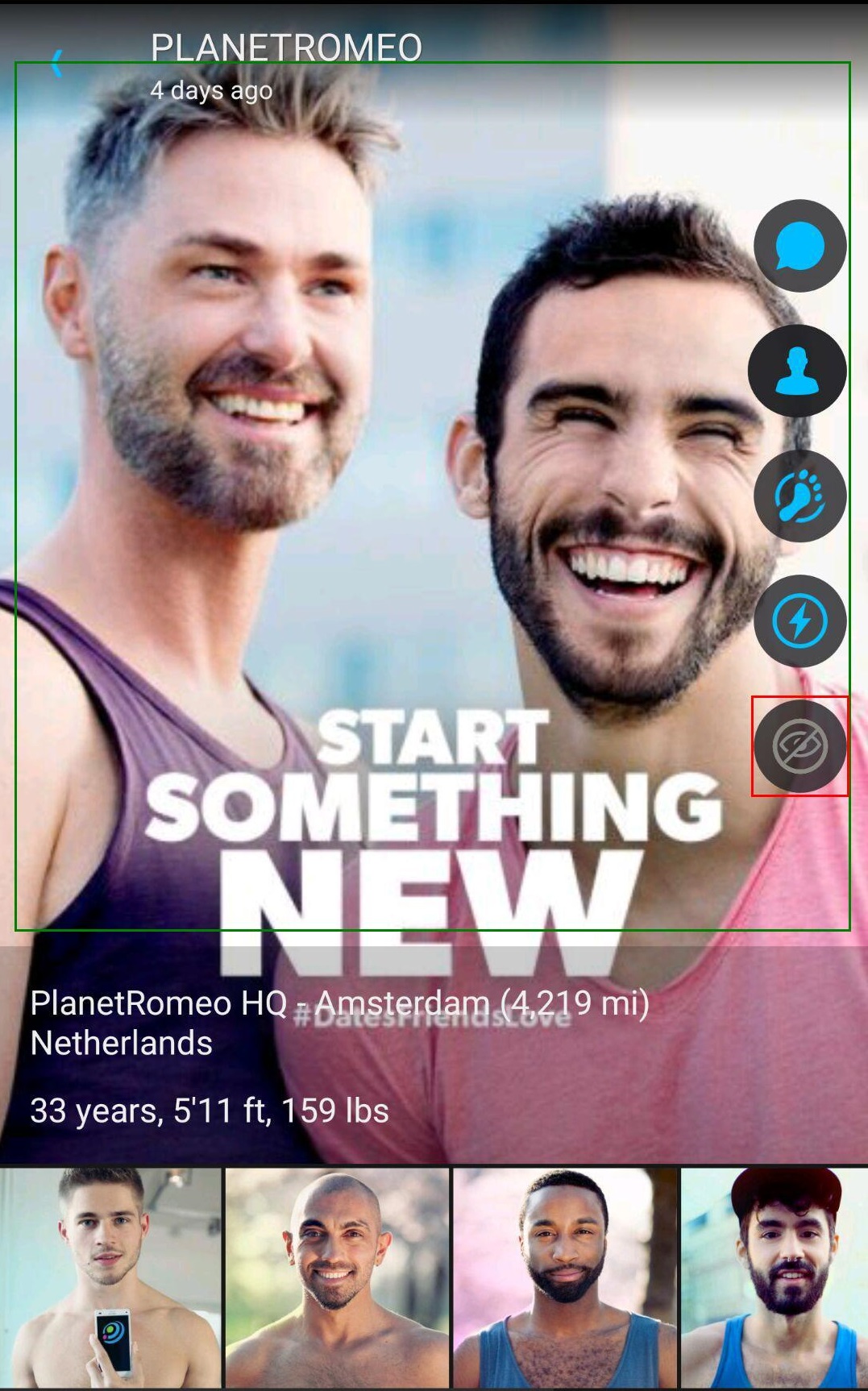}
}
\hfill
\subcaptionbox{
NYC Subway Time:
``go to step 3'', ``third bus stop'', ``view alternate route'' (ref);
``select number 3'' (PG);
``select line 3'' (\toolName{}).
\label{fig:screen_good_train}}{
    \includegraphics[width=0.23\textwidth]{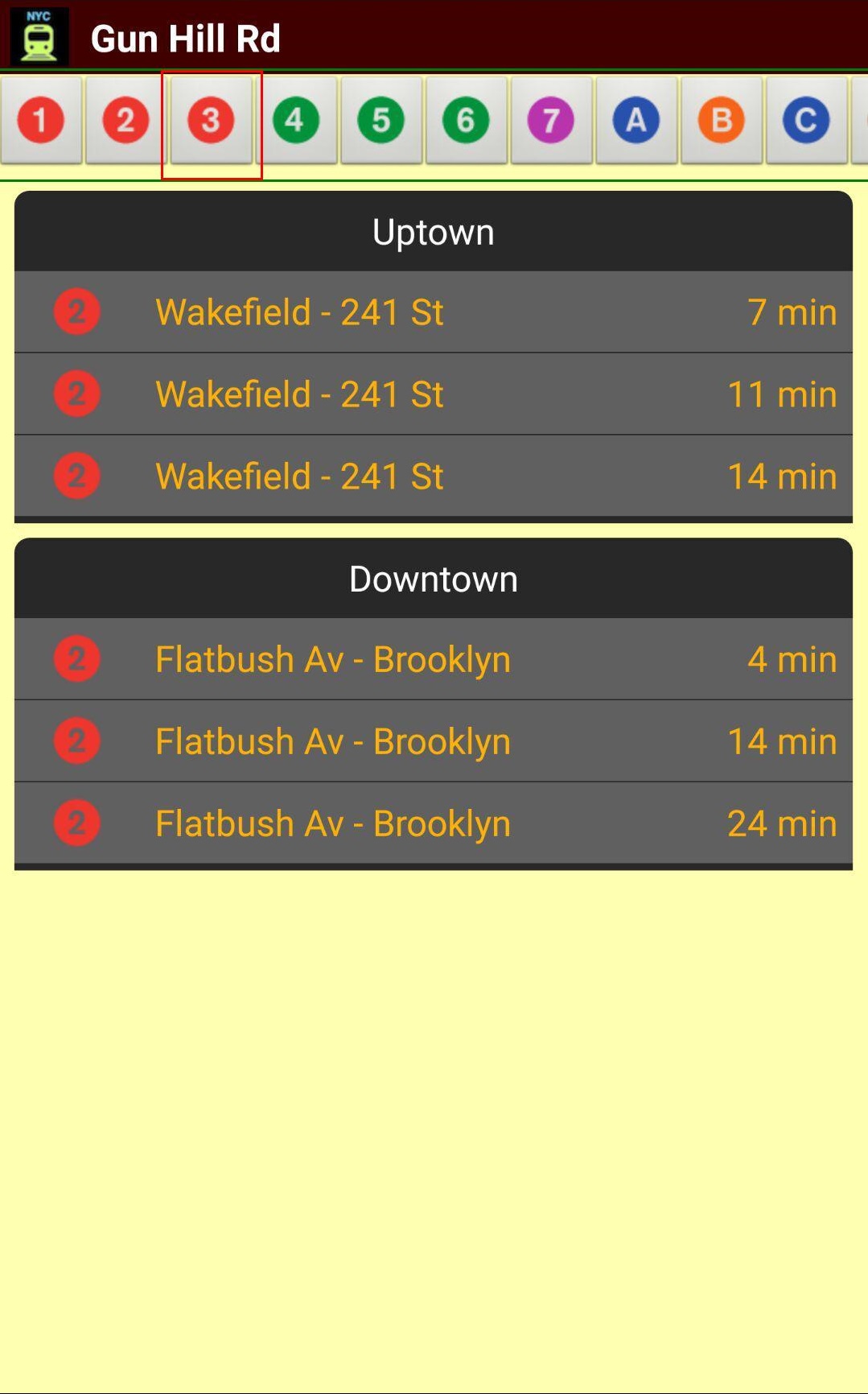}
}
\hfill
\subcaptionbox{
Tablet Calling:
``select angola'', ``select angola'', ``set your location to angola'' (ref); 
``flag'' (PG);
``select angola'' (\toolName{}).
\label{fig:screen_good_angola}}{
    \includegraphics[width=0.23\textwidth]{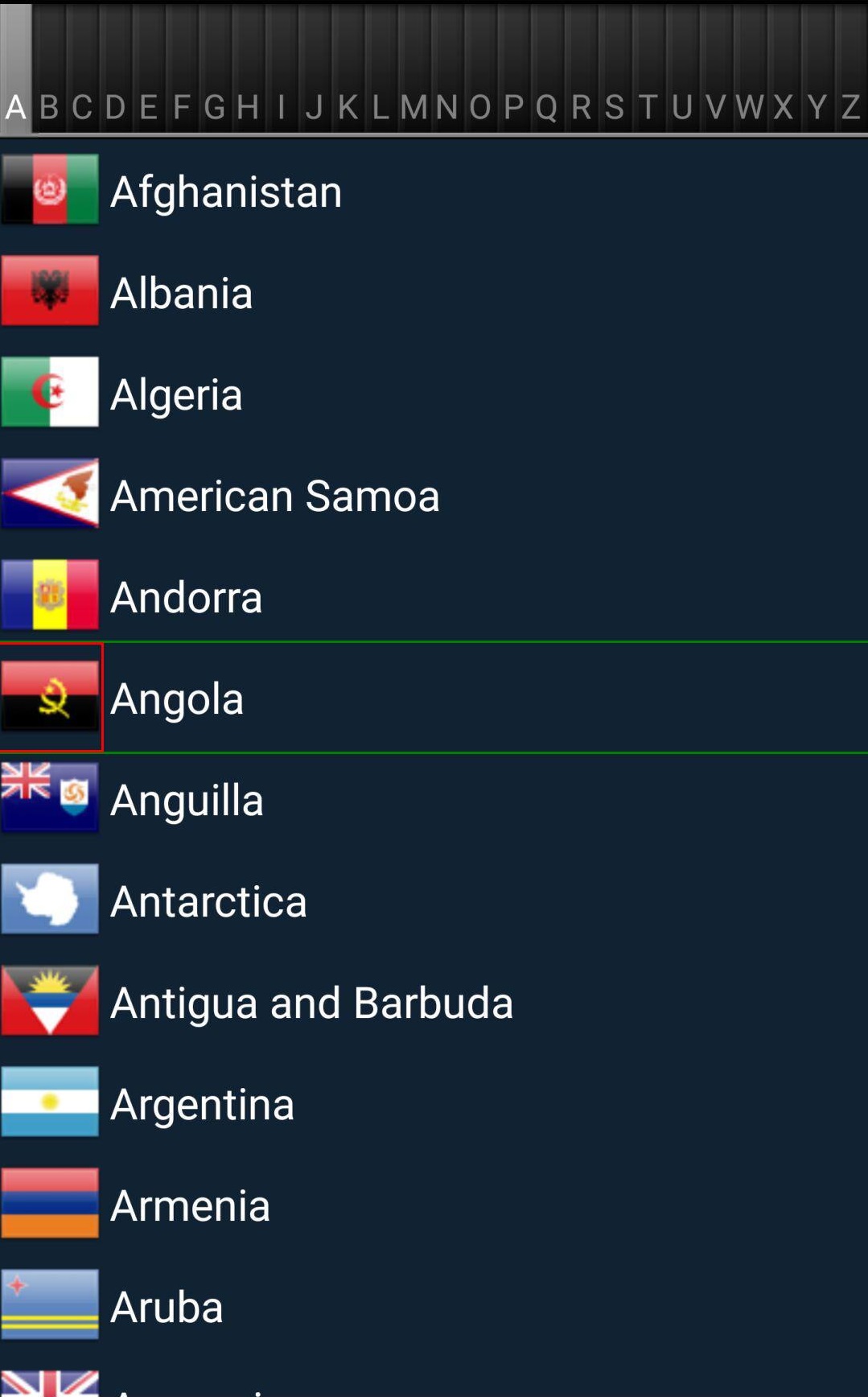}
}
\caption{Sample double-anonymous user survey screens (RQ4):
target icon (red box),
icon's parent (green box, not marked in survey or tool input),
app name:
ground truth (ref),
PaliGemma-448\textsubscript{c} (PG),
\toolName{}-TextT \& \toolName{}-MMT\textsubscript{i} (\toolName{}).
}        
\label{fig:good_examples}
\end{figure*}

\subsection{RQ3: Ablation Study}

Table~\ref{tab:contribution_of_inputs} explores the contribution of \toolName{}'s components. We start with the two best-performing variants (\toolName{}-TextT and \toolName{}-MMT\textsubscript{i}) fine-tuned on their respective full training sets. We then measure the impact of removing select elements from their Listing~\ref{lst:prompt_TextT} and~\ref{lst:prompt_MMT} prompts via our two main automated metrics.

\begin{table}[h!t]
\centering    
\caption{Comparison of fine-tuned \toolName{}'s performance after removing certain components from input.
}
\label{tab:contribution_of_inputs}
\begin{tabular}{llrr}
\toprule
\toolName{} & Input & CIDEr & SPICE \\
\midrule
TextT & all & 134.3 & 22.3 \\
TextT & w/o in-icon text (OCR) & 130.5 & 21.9 \\
TextT & w/o icon's resource-id & 129.3 & 21.2 \\
TextT & w/o parent \& sibling DOM info & 125.7 & 20.1 \\
\midrule
MMT\textsubscript{i} & all input & 138.3 & 23.2 \\
MMT\textsubscript{i} & w/o in-icon text (OCR) & 136.6 & 22.3\\
MMT\textsubscript{i} & w/o icon's resource-id & 135.0 & 23.0 \\
MMT\textsubscript{i} & w/o parent \& sibling DOM info & 130.4 & 20.5 \\
\bottomrule
\end{tabular}
\end{table}

At a high level, these experiments are consistent with the overall results, i.e., \toolName{}-MMT\textsubscript{i} scores higher than \toolName{}-TextT in every configuration across both metrics. We observe the largest performance drop from removing parent and sibling DOM info, by some 8 CIDEr points for both variants. This underlines the value that local DOM context provides to alt-text generation. 

With some distance the next-biggest impact has the icon's DOM resource id. This impact was larger for the variant fine-tuned only on text (5 CIDEr point drop for \toolName{}-TextT vs. some 3 for \toolName{}-MMT\textsubscript{i}), showing that the icon's resource id encodes some of the icon image's in-context meaning. Finally, a similar observation applies to the OCR-inferred in-icon text. Removing this information leads to a larger CIDEr drop in \toolName{}-TextT (by some 4 points) vs. in \toolName{}-MMT\textsubscript{i} (by some 2). At the same time, OCR-inferred in-icon text has the smallest impact of the different components.

\subsection{RQ4: Humans Rate \toolName{} Results Highest}

To explore what the automated metric results mean in human terms, we also conducted a small double-anonymous online survey to assess how human evaluators rate the generated alt-texts' quality. Specifically, we want to know how human raters score both the ground truth and the alt-texts generated by the models that scored highest on CIDEr and SPICE on partial screens, i.e., 
PaliGemma-448\textsubscript{c}, 
\toolName{}-TextT, and 
\toolName{}-MMT\textsubscript{i}.

The survey contains 50 randomly selected WC20 test screen/icon pairs (icon marked via red bounding box). For each pair, the survey also presents the app's name and four (anonymized) alt-text options in a randomized order (3 tool outputs plus one randomly selected human-written ground truth label).
The survey instructs participants to review each app screen, understand the four alt-text descriptions in the context of helping ``visually impaired users understand the functionality and purpose of the highlighted icon'' (accurately, clearly, and concisely), and rate each of the four alt-texts on its ``appropriateness and meaningfulness in the screen's context'' on a scale from 1 (``very poor'') to 5 (``excellent'').

As smartphones are ubiquitous we sampled one author and three (non-author) students (two without a computer science background) as participants, yielding 800 total ratings. On average, participants rated the alt-text of the ground truth~4.1, PaliGemma\textsubscript{c}~3.6, \toolName{}-TextT~3.9, and \toolName{}-MMT\textsubscript{i}~4.1. These results are consistent with our Table~\ref{tab:results_metrics} automated metric scores, i.e., (1)~both \toolName{} variants rate higher than PaliGemma\textsubscript{c} and (2)~\toolName{}-MMT\textsubscript{i} rates higher than \toolName{}-TextT. Maybe most interestingly, human annotators rated \toolName{}-MMT\textsubscript{i} on par with the human-written ground truth.

Rating variability, expressed via standard deviation~(SD), for PaliGemma\textsubscript{c} and \toolName{}-TextT (SD=1.4 each) was higher than for \toolName{}-MMT\textsubscript{i} (SD=1.2), suggesting \toolName{}-MMT\textsubscript{i} has a slightly more consistent reception across evaluations.

\begin{table}[h!t]
\centering
\caption{Hypothesis testing via Wilcoxon signed-rank test.}
\begin{tabular}{lcc}
\toprule
Comparison & p-value & Reject null hyp. \\
\midrule
\toolName{}-MMT\textsubscript{i} vs PaliGemma\textsubscript{c} & $7.23 \times 10^{-7}$ & Yes \\
\toolName{}-TextT vs PaliGemma\textsubscript{c} & $4.5 \times 10^{-3}$ & Yes \\
\toolName{}-MMT\textsubscript{i} vs ground truth & 0.47 & No \\
\toolName{}-TextT vs ground truth & 0.39 & No \\
\bottomrule
\end{tabular}
\label{tab:wilcoxon_rank}
\end{table}

Given the failure of the Shapiro-Wilk test for normality~\cite{Shapiro-wilk}, we opted for the Wilcoxon signed-rank test~\cite{wilcoxonranktest}, a non-parametric method suitable for our samples. The Table~\ref{tab:wilcoxon_rank} results indicate that both \toolName{}-MMT\textsubscript{i} ($p < 0.0001$) and \toolName{}-TextT ($p = 0.0045$) significantly outperform PaliGemma\textsubscript{c}. Furthermore, there is no significant difference between \toolName{} and the ground truth.

Figure~\ref{fig:good_examples} shows 4/50 survey screens \toolName{} performed well on. We make the following observation. 
First, ground truth labels sometimes conflict with each other (as in Figure~\ref{fig:screen_good_train}), as WC20 annotators were only given screenshot, icon bounding box, and a few sentences worth of app description from the Google Play Store---but not access to app internals. 
Second, \toolName{} benefits from the DOM tree, e.g., by extracting the substrings ``Backspace'', ``hide'', and ``line3'' from the icon's DOM node in the Figure~\ref{fig:screen_good_calc}, \ref{fig:screen_good_romeo}, and~\ref{fig:screen_good_train} examples and ``Angola'' from the icon's lone sibling node (Figure~\ref{fig:screen_good_angola}). 
Finally, extracting image text does not compensate for extracting DOM information, both for in-icon text (``3'' in Figure~\ref{fig:screen_good_train}) and sibling text (``start something'' in Figure~\ref{fig:screen_good_romeo}).

\subsection{RQ5: User Survey on \toolName{} Prototype}

To get feedback on practitioners’ perceptions of \toolName{}, we conducted an anonymous online survey among software developers. The survey first asks participants to watch a 4-minute demo video (included in \toolName{}'s replication package~\cite{figshare}) of \toolName{} used as an Android Studio plugin. We distributed this survey on Google Forms via multiple channels, including LinkedIn, Facebook, and Reddit, receiving 24 responses.

Each participant self-identified their roles. Most common were software engineer (54\%), mobile app developer (50\%), and UI/UX designer (21\%). Mobile app development experience varied, with 22\% under 1 year, 52\% 1--3 years, 17\% 4--6 years, and 9\% 7+ years. The most popular primary development platforms selections were Android (71\%), followed by web (25\%), cross-platform (React, etc.) (21\%), and iOS (13\%). 

The majority of participant rarely (29\%) or never (25\%) adds alt-text to UI elements. 33\% add alt-text when required and 13\% add alt-text to all UI elements. Only 17\% report having used a tool to generate alt-text (``\emph{Have you tried any other approach to generate alt-text automatically (e.g., another IDE plugin, upload icon to Google, OpenAI, etc)}''). 
When asked about the demo video, 88\% of respondents found it sufficient (``\emph{Did the demo video clearly explain how the plugin works?''}). On a 1--5 star scale, participants rated \toolName{}'s perceived usefulness~4.2 (SD=0.8). On the same scale, participants rated their likelihood of recommending the plugin to a colleague or friend who develops mobile apps~3.9 (SD=1.3), with 71\% rating it favorably ($\geq$4). Two out of three participants selected that they would use the plugin regularly (30\%) or occasionally (35\%). A further 22\% would use it if improvements are made and 13\% would not use it.

Participants also provided suggestions for improvement, including supporting other UI element types, major IDEs, and Jetpack Compose (\textit{``its good to go. But now i am using compose for native android view. so if you could do it for compose it will be great thing.''}) We consider these suggestions as relatively straight-forward engineering additions.

\subsection{Causes of Low RQ1 Metric Scores}

\begin{figure*}[h!t]
\centering
\subcaptionbox{NYC Subway Time:
``first page'', ``open selected window'' (ref); 
``select number'' (PG);
``select line'' (TextT); 
``select line 1'' (MMT).
\label{fig:context_effect}}{
    \includegraphics[width=0.23\textwidth]{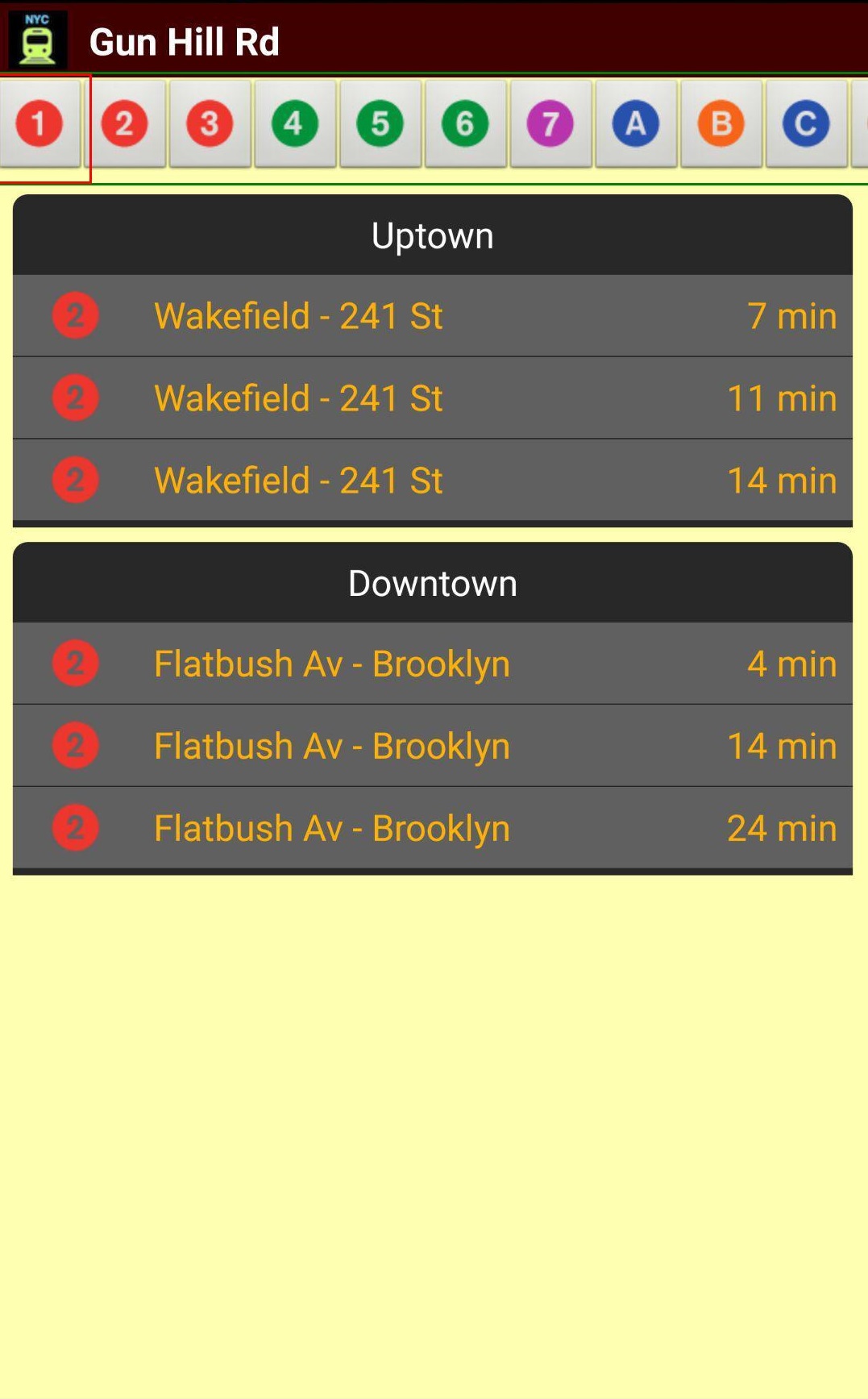}
}
\hfill
\subcaptionbox{Boat Browser Mini:
``advertizer description'', ``click on advertisement'', ``this is an advertisement'' (ref); 
``play video'' (PG);
``play button'' (\toolName{}); 
\label{fig:image_effect}}{
    \includegraphics[width=0.23\textwidth]{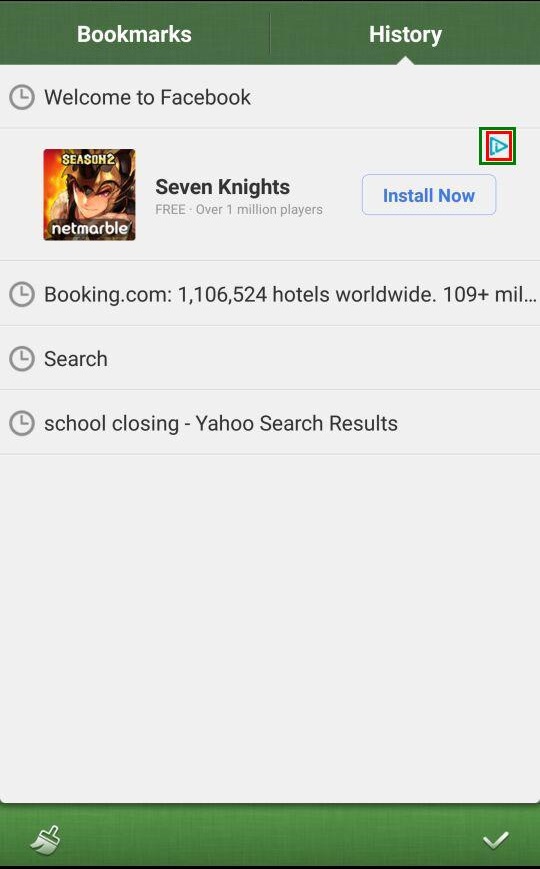}
}
\hfill
\subcaptionbox{Tip N Split Tip Calculator:
``increase button'', ``increase count'', ``plus'' (ref);
``add more''; (PG)
``add person'' (\toolName{}).
\label{fig:valid_nonmatching}}{
    \includegraphics[width=0.23\textwidth]{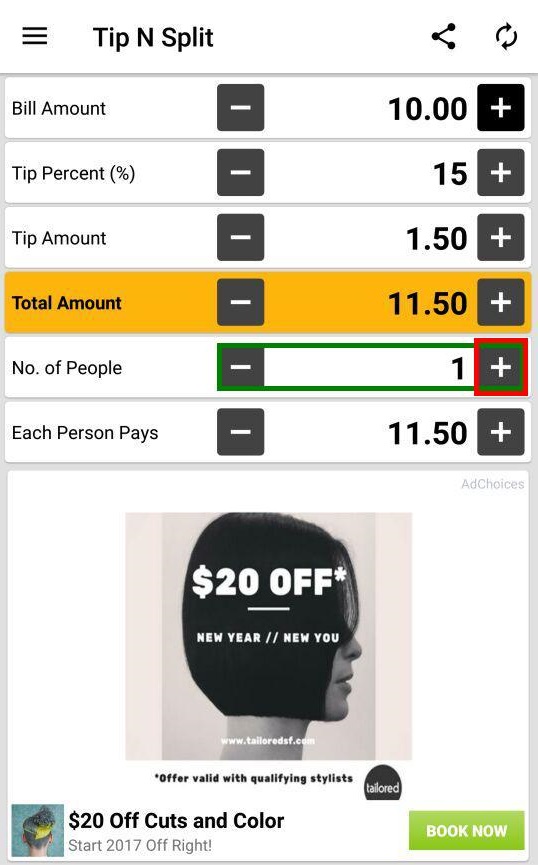}
}
\hfill
\subcaptionbox{Hi:
``refresh button'', ``share post'' (ref); 
``delete'' (PG);
``select privacy'' (\toolName{}).
\label{fig:vague}}{
    \includegraphics[width=0.23\textwidth]{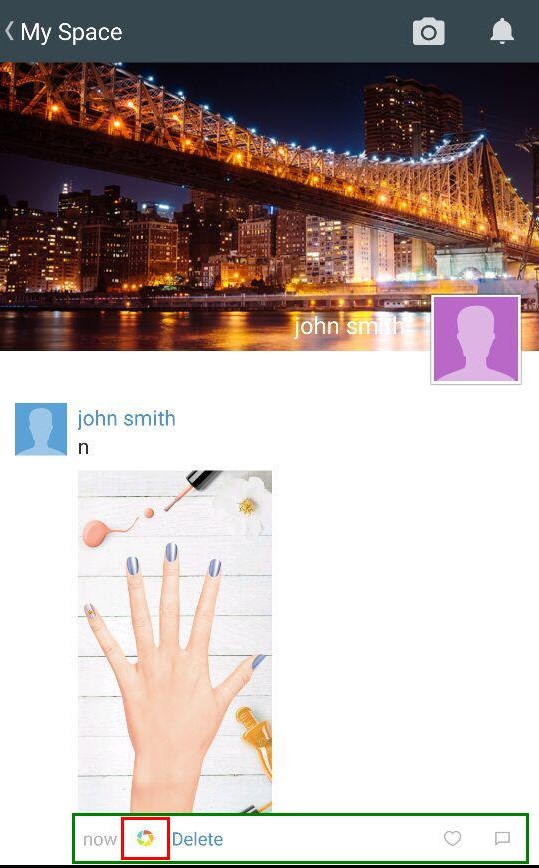}
}
\caption{Screens where \toolName{} infers sub-optimal alt-text;
target icon (red box),
icon's parent (green box, not marked for tools),
app name:
ground truth (ref),
PaliGemma-448\textsubscript{c} (PG),
\toolName{}-TextT (TextT),
\toolName{}-MMT\textsubscript{i} (MMT),
TextT \& MMT (\toolName{}).
}
\label{fig:failed_cases}
\end{figure*}

To better understand \toolName{}'s limitations, we manually review random test screens on which \toolName{} scored low on automated metrics (RQ1). Specifically, we follow WC20 and sample 50 of the around 200 test icons for which \toolName{}'s alt-texts have the fewest overlaps with ground truth labels. Following are the four main causes we identified, several of which may affect a given icon.

\subsubsection{Icon Context Misleads Inference (16/50)}

In several cases the DOM tree information \toolName{} extracts misleads alt-text inference. In the (extreme) Figure~\ref{fig:context_effect} example, the DOM tree info of both the icon and its sibling nodes contain the substring ``line'' a total of 25 times. While the resulting \toolName{} alt-text deviates from the ground truth, in this case \toolName{}-MMT\textsubscript{i} still produces meaningful alt-text.

\subsubsection{Icon Image Misleads Inference (11/50)}

In some cases, taking the icon image literally misleads \toolName{}. For example, the Figure~\ref{fig:image_effect} icon resembles both a stylized Google Play Store icon and the widely used media ``play'' icon. \toolName{}-TextT's icon-only label is ``Play button icon''. The DOM tree lacks the icon's resource id but contains the substring ``AdChoices'' both in the sibling and parent. Given that the button is part of an ad to install another app from the Play Store, \toolName{}'s alt-text is relatively close but does not fully capture the button's meaning.

\subsubsection{Non-matching But Valid Alt-texts (10/50)}

In some cases, \toolName{}-generated alt-text is semantically correct but describes the icon differently than the ground truth labels. For example, in Figure~\ref{fig:valid_nonmatching} the DOM contains resource id ``no\_people\_up'' for the icon and ``no\_people\_down'' for a sibling. The resulting \toolName{} label is meaningful but quite different from the ground truth.

\subsubsection{Vague Icons (7/50)}

In what we call a vague icon the ground truth labels differ widely from each other, indicating that human annotators where confused. In the Figure~\ref{fig:vague} example, the icon's DOM tree resource id is ``privacy\_icon'', which conflicts with both ground truth labels and to us also does not appear to be an intuitive meaning for this icon image. In such cases adding DOM information can be crucial to generate meaningful alt-text (even though in this case the automated metrics penalize \toolName{} for not matching the apparently confused ground truth).

\section{Related Work}

Most existing accessibility tools operate post-development, requiring developers to manually revise UI elements after accessibility violations are detected. This reactive approach increases technical debt and often results in accessibility issues persisting in production. In contrast, \toolName{} integrates alt-text generation into the IDE, ensuring that accessibility is addressed early during UI development. 

\subsection{Multi-modal Models for UI Understanding}

Large multi-modal models have significantly improved UI understanding, enabling automated screen comprehension, interaction modeling, and accessibility enhancements. Prior work has explored their ability to process view hierarchies and screen layouts to enhance usability and automation. Wang et al.~\cite{wang2023enabling} demonstrated how LLMs summarize mobile screens and assist with navigation, while AutoDroid~\cite{wen2024AutoDroid} combined commonsense knowledge with automated app analysis~\cite{wen2024AutoDroid} to enhance task automation. Beyond UI comprehension, LLMs have been used for accessibility testing. AXNav~\cite{taeb2024axnav} integrates LLMs with pixel-based UI analysis to execute accessibility tests via natural language instructions. ILuvUI~\cite{jiang2023iluvui} fine-tunes LLAVA by using synthetic UI datasets to improve UI specific tasks. AMEX~\cite{AMEX_2024} employs GPT-4o to generate descriptions of UI elements and screen layouts, providing structured UI metadata for automated GUI-control agents, while ScreenAI~\cite{baechler2024screenai} enhances UI parsing and widget interpretation through a vision-only approach. While these works excel in screen-level UI understanding, they focus on full-screen interactions rather than the partial UIs available during development.

\subsection{Enhancing Mobile UI Accessibility}

Accessibility issues in mobile apps, such as missing item labels, vague descriptions, and poor contrast, create significant barriers for visually impaired users ~\cite{chen2021accessible}. Despite accessibility guidelines, many apps fail to provide adequate alt-text, limiting the effectiveness of screen readers~\cite{fok2022large, ross2018examining}.
To address these challenges, researchers have explored automated accessibility enhancements. Chen et al. proposed a comprehensive labeling method for icons, combining image classification and few-shot learning~\cite{chen2022towards}. Zang et al. introduced a deep learning-based multi-modal approach that combines pixel and view hierarchy features to improve icon annotation in mobile applications~\cite {zang2021multimodal}, enhancing icon detection and classification. 
Beyond icons, accessibility gaps extend to other UI components. Missing hint-text in form fields can hinder usability for screen reader users, prompting Liu et al.~\cite{liu2024unblind} to develop an LLM-based method for generating meaningful text input hints. Similarly, poor color contrast reduces text readability, leading Zhang et al.~\cite{zhang2023automated} to introduce Iris, a tool that dynamically adjusts color settings in mobile apps to ensure compliance with accessibility standards. AccessiText identifies accessibility violations related to text readability, including contrast issues and insufficient font size settings~\cite{alshayban2022accessitext}. ScaleFix automates UI scaling repairs, preventing text and display settings from distorting interface layouts~\cite{alotaibi2023scalefix}.

\subsection{AI-assisted Development in IDEs}

Modern IDEs incorporate AI-driven tools to automate repetitive tasks, improve code quality, and provide intelligent recommendations. While LLM-powered coding assistants improve code completion and debugging, accessibility support remains largely underexplored.
Recent research explores AI-driven IDE enhancements, such as IDECoder, which integrates IDE-derived context for better code completion~\cite{li2024enhancing}. Zharov et al. explore LLMs as universal interfaces for automating multi-step IDE tasks~\cite{zharov2024tool}. Specialized plugins like Coconut detect privacy violations in Android Studio~\cite{li2018coconut}. Adee, a plugin for Figma and Sketch, enables real-time accessibility evaluation by providing a touch target size checker, contrast analysis, and an alt-text generator~\cite{hadadi2021adee}.

AI coding assistants like GitHub Copilot have the potential to improve accessibility, but developers' expertise and intervention on the matter are needed. However, developers often prioritize efficiency over accessibility unless guided by explicit feedback~\cite{mowar2024tab}. 

IDE support for generating meaningful descriptions during app development remains limited. Android Lint detects contentDescription attributes but does not assist in generating meaningful alt-text, leaving the burden entirely on developers~\cite{lint}. AI coding assistants, such as GitHub Copilot, can improve accessibility but do not consistently generate accessible code unless explicitly prompted by developers. This introduces a critical dependency on developer awareness~\cite{mowar2024tab}. Even when prompted, Copilot-generated code often requires manual refinement, as it may include placeholder alt-text that developers must manually update. Recent efforts to address this issue include CodeA11y, a GitHub Copilot extension that reinforces accessible development practices throughout the coding process~\cite{mowar2025codea11y}. CodeA11y enhances accessibility awareness by generating UI code that aligns with accessibility standards, prompting developers to address existing errors and reminding them to complete any placeholders. However, while these reminders improve compliance, developers must still write the alt-text themselves, making the process manual and time-consuming.

\section{Conclusions}

Alt-text is essential for mobile app accessibility, yet UI icons often lack meaningful descriptions, limiting accessibility for screen reader users. Existing approaches either require extensive labeled datasets, struggle with partial UI contexts, or operate post-development, increasing technical debt. We first conducted a formative study to determine when and how developers prefer to generate icon alt-text. We then explored the \toolName{} approach for generating alt-text for UI icons during development using two fine-tuned models: a text-only large language model that processes extracted UI metadata and a multi-modal model that jointly analyzes icon images and textual context. To improve accuracy, the method extracts relevant UI information from the DOM tree, retrieves in-icon text via OCR, and applies structured prompts for alt-text generation. Our empirical evaluation with the most closely related deep-learning and vision-language models showed that \toolName{} generates alt-text that is of higher quality while not requiring a full-screen input.

\bibliographystyle{ACM-Reference-Format}
\bibliography{ref}
\nocite{*}

\end{document}